\documentstyle[twocolumn]{article}
\title{New Features in Non-coalescense of Tangent Liquid Bulks}
\author{Sanli E. Faez \footnote{\it Email:Sanli@aban.ysc.ac.ir} \\ \it Sharif University of Technology\\ \it Young Scholars Club}
\date{}
\begin{document}
\maketitle
\begin{abstract}
We had observed a couple of new phenomena in which, two liquid bulks 
in contact with each other do not coalesce. The main reason seems to be
the motion of the surface in the place of contact, which forms an air film
between the bulk boundaries. The surface velocities is also estimated and
showed to be near each other and also near the surface velosities
in the similar phenomena reported before. 
\end{abstract}
\section{Introdution}
When two drops of one liquid come in contact,
they simply coalesce and form a single drop.
This is to minimize the free surface of the liquid, and accordingly minimize
the surface potential energy caused by surface tension\cite{Eggers}.
But can we set the circumstances such that the two tangent drops
do not coalesce, even though externally pushing them toward each other?
The answer is yes, and this is the main point of discussion in this article.

We have recently found a class of effects in which two
liquid bulks of the same material touch each other,
but do not coalesce. In all cases there is even an additional force
pushing the bulks toward each other.

These phenomena can be divided into two categories. The first category is
the floating of a liquid drop on the free surface of the same liquid.
This free surface is, sometimes, the flat surface of the
liquid in a container and sometimes the surface of another drop.
This phenomenon had been seen before in two different cases;
we have observed a third case.

In the first case a traveling water droplet floats on the surface of the
water, till it is moving faster than some critical velocity
\cite{Shul, Reynolds}.

The second case is completely different in appearance:
Two liquid drops of the same kind but with different temperatures are
pushed to each other but do not coalesce \cite{monti}, because the tempreture gradiant
causes a surface tension gradient, which leads to the
formation of a film of air between the drop surfaces \cite{Del}.

The third case, is the floating of a droplet of water on the surface
of water (fig 1). This droplet is standing still with respect to
the container, but there is a wave on the surface of the water
which causes the non-coalescance\cite{peyman}.

The second category is mainly based on the currents on surface of a liquid.
The main point is that two currents, which may have slightly or completely
different directions, collide with each other. Under some circumstances,
instead of combining, the currents repel each
other just like the collision of two solid balls; see fig 2.

In the following, we will introduce the effects in some detail
and describe some observations on the phenomena.

\section{The Phenomena}
\subsection{Floating of droplets over the
oscillating surface of water}
This phenomenon can simply be seen in a cylinderical plastic bottle.
of about 30cm in height and 10 cm in diameter.
If you pour
water in the bottle for about three-forths of the bottle's volume, and
then strike the bottle repeatedly with your finger, a little below the surface
of the water, you will see some droplets of different diameters (1-3mm).
These droplets come from the walls, go to the center
and stay there still for some seconds.
They only float if you continue striking, and sink immediately if the
surface stops oscillating. You can improve the number and
life time of the drops up to minutes by changing the period of striking.
More success will be gained by a little more practice.
These are not air bubbles. They are much brighter,
and if you blow at the surface, they move much easier and faster than
bubbles. You can see that their
buttom is placed below the water surface, while in the case of
bubbles they are completely over the surface. This shows that they are
much heavier than air. What else can they be but just water?

If you make more floating drops they will come together, and
sometimes make big colonies of about fifty in number (fig 3). This colonization
can be explained easily if we mention that each droplet deforms the
surface of the water around it like a hole. If two droplets get near
each other they slide toward each other and make a unique deeper hole
that lets the droplets go down more.
If you stop hitting the bottle, the drops coalesce into each other
as time passes, and form bigger and bigger droplets, up to diameters
of about 8mm, till they all sink. In these colonies, you can see
the rainbow if you watch carefully.
\subsection{Repulsion of two colliding water currents}
If we bring two narrow cylindrical currents of
water (about 3mm in diameter){\em slowly} in contact with each other, {\em sometimes} instead of
combining into one current, they just repel each other (fig 2),
like the case of collision between too hard bodies.
This state is stable for
minutes, and it becomes more stable if we fix the initial conditions of
the currents more precisely. We have dyed one stream of water and observed
that the streams do not combine at all.

Each current deforms slightly {\em before} the point of
collision, so we can see some kind of wave on the surface of the currents.
After the collision point, the shape of water changes its form completely.
We see that the cross section of the current starts changing its shape from
circular to elliptical and viceversa, periodically, if we travel along the length
of the flow. This is commonly expected because the circular cross section
is just the case of lowest potential and with any change in the initial
cross section we must expect oscillation of the shape.

This phenomenon is not only for cylindrical currents. It is seen in the case
of other kinds of collision, for example a droplet that hits the surface
of the water in a close angle and then jumps up from it.
In another experiment
we tested this for collision between a cylindrical current and a plain one.
(See fig 4).

To make the plain current, we used the scattering of a falling current
of water from the surface of a spoon.
You might have experienced this, especially if you are not an expert in
washing dishes . In this case again, you can see some kind of deformation
just like the case where both currents were cylindrical.

\section{The Experiments}
\subsection{Floating Droplets}
The experiments with the droplets first began with the simple setup
just described in the previous part. The first step was to make some
machine that would do work of a human. This machine made the work much easier
so that we could fix the frequency of the strikes, and change it to any number
we wanted. The most important result of this
experiment was that there is a set of distinct frequencies that let droplets
{\em float} more, both in number and lifetime, and there is a
different set of frequencies in which
droplets are {\em created} much more than other frequencies.
These two different
sets showed that creation and floating of the droplets are two completely
distinct effects. We can simply float drops from an external source;
for example a faucet. The size of the initial drops (before combining with
other drops) depends mainly on the wave spreads around the plastic bottle,
that is, the \underline{sound} of strike. The result: Drops made from bass
strikes are bigger than drops made from treble one.

A very strange observation was that adding a little detergent to water
allows the drops to construct and to float more. This behavior might be
caused by the special form of detergent molocules, with two defferent ends.

The second step was to change the shape of the container. We chose
a square one, so that we can analyse the wave spreading over the surface,
easier. The main change was the set of frequencies. Floating of drops
was just like before.

Next, we tried to analyze the movements inside the drop.
We used ink, since we could easily inject it to the drop (fig 5).

In this way we can even make the drops
bigger or smaller by injecting water into them, or sucking out of them.
The result was that the velocities in the drop are very small in
compare with the velocities on the free surface, because the ink
defuses much
slower in the drop than the surface of the water. It means that the water in the
drop is somehow motionless. The ink in the drops can not be spread to the below
water until the hole body of drop sinks (coalesces).
This is just like the two flows mentioned in the previous part.

Other experiments that have been done are:
\begin{enumerate}
\item
If we look from below the surface, we'll see the bright
surface of drops, this shows that there is something else with different
refraction coefficient from water between the two water layers,
(otherwise, we couldn't see the layer) and what else can it be but air.
\item
We can separate one drop from others, by enclosing it with a ring. 
This ring do not allow the drop to travel freely over the surface
and also preserves it from others to come in contact with. By this trick we can
improve the lifetime of drops very much.
\item
We can make the air pressure higher and see a great improve
in the number and lifetime of the drops. The experiment is done in 
a closed plastic bottle of cola, which had been shaked before the 
experiment began, and in result had a high pressure inside.
We can easily decrease the pressure and see the phenomenon vanishes.

\item
Some drops do not sink at once. These drops loose their mass,
and make a smaller floating drop. This can be repeated two
or three times. An existing explanation is that this is due to
the existance of impurities on the free surface of water, \cite{Reynolds}.
If the impurities are larger than
the air gap between the two layer, and because water wets the impurities,
a bridge between the bodies form that transfers water from the droplet
to the free surface. Sometimes the
flow caused by this transferation shoots the impurity out and we gain
a new smaller floating droplet. The main point here is wetting quality
of water. We know that in a non-wetting liquid, for example mercury,
the impurities on the surface, do not let the drops to coalesce.
\end{enumerate}

\subsection{Colliding Flows}
In the previous section we mentioned how we can see this effect.
here we bring some other experiments related to the phenomenon.
Experiments show that, if the velocity of the current
is less than, or more than some critical values, this
phenomenon can hardly be seen.
The lower limit is, roughly, few centimeters per second, but
we couldn't found any fixed value, because it depended
hardly on the circumstances. We can just say that {\em there is} some lower
limit. There is also an upper limit for the velocities, about few
hundred centimeters per second, though it is not very sharp limit.
In fact, the effect can be seen even in higher velocities, but it becomes very
sensitive and you should fix the initial conditions much more precisely,
to obtain a mentionable stability lifetime.

In all the related cases, we see that there are some kind of
movements in the surface of liquid bulks. So we can agree with the
idea that water drags air between the two layers, and the film of
air prevents two surfaces from reaching each other and coalesceing.
Though not expected, dimensional analysis shows that the order
of the velocities, in all the four cases, are the same.

\section{Dimensional Analysis}
If we want to compare this new effects with the effects that was
\cite{Shul, Del}
we should first estimate some parameters of those experiment.
As said before, the most important parameter is
the surface velocity of two neighbor layers. we call this velocity
$v_s$.  In the second place we can put the force, $f$, pushing two bulks
towards each other. We can then estimate the thickness, $t$ ,of the
air layer and the pressure that it can support before it ruins.

\subsection{Surface velocities}
For the traveling droplets described in the introduction, we can take
the drop speed as the surface velocity in the contact point. As described
in the article\cite{Shul} the drops are produced from a jet that
throws them in a 45 degree initial angle, And the drops hit the surface
just at their maximum height.
Because of the specific initial projection, the horizontal speed
equals the initial vertical speed. This means
$$ v_s=v_{0x}=v_{0y}=(2gh)^{1/2} $$
where $h$ is the maximum height of the drops trajectory.
If we take $h=.05m$, the estimated answer is $$v_s=1 m/s$$.

The next experiment is non-coalescense of two drops with different
temperatures. For this part we should first estimate another parameter,
the velocity correlation length, $l_c$, caused by viscosity. We estimate it
by pure dimensional analysis. as we know the most useful dimensionless
number in this kinds of problems is the {\em Reynolds Number}.
We also know that this number in cases - for example a rigid body in a flow -
can be calculated as $\rho v/\eta l$, where $\rho$, $\eta$ and $v$ are
density, viscosity and velocity of the flowing liquid, and $l$ is a parameter
with length dimension, related to the size and shape of the rigid body.
So we can take $\eta/\rho v$ as, $l_c$.

Now we go back to calculation of $v_s$. We know that here the cause of
surface motion is the temperature gradient that causes gradient of surface
tension. We can write
$$ \frac{d\sigma}{dx}=\frac{d\sigma}{dT}\cdot \frac{dT}{dx} $$
Here $\sigma $ and $T$ are surface tension and temperature.
If we calculate the net force on a surface differential area
we find that $\Delta f=w \Delta x d\sigma/dx$, where $w$ is the width,
and $\Delta x$ is the length of the element. Now we consider that
this force should move the volume $w\Delta x t$. Here $t$ is the
typical thickness of the moving layer, and we put $l_c$ in its place.
So we can write a simple Newton Equation and obtain that
$$ a_s=\frac{d\sigma}{dx}\frac{1}{\rho l_c} $$
and $a_s$ is the surface acceleration. If we take the diameter of
the drop $d$, as an estimate for the length in which the surface material
accelerates, and also put the quantity of $l_c$ in its place,
we find that $$ v_s=(\frac{d\sigma}{dx} \frac{v_s d}{\eta})^{1/2} $$
Solving the equation for $v_s$ and putting in place of $d\sigma/dx$
we find the last equation:
$$v_s=(\frac{d\sigma}{dT})\frac{\Delta T}{\eta}$$
On the basis of  data given in \cite{Del} we take the parameters
like this: $\eta=5*10^{-3}kg/ms$, $\Delta T=10 K$ and
$d\sigma/dT=5*10^{-4}m/{s^2 K}$. The result estimates that
$$v_s=1m/s$$. The result is strangely like the previous case. This is a good
win for the idea that this effects has the same physics. And also
courages us to continue this analysis for the other phenomena.

For the other floating drops reported in this article we can easily
estimate $v_s$ if we take the oscillation of water a gravity wave with
small amplitude. The answer is $v_s=(A/\lambda)(2\pi g\lambda)^{1/2}$.
Where $\lambda$ is the wavelength and $A$  is the amplitude.
By taking $\lambda=2mm$ and $(A/\lambda)=.5$ we obtain,
$$v_s=.2 m/s$$
Again a near answer to the two previous.

And at last for colliding currents. Because the motion of each current is a simple
free fall, we can take $v_x << v_y $. And so $v_s=v_y=(2gH)^{1/2}$.
Where $H$ is the vertical distance between the falling point of water
and the collision point. Again we take $H=.05m$ and another time we find
$$v_s=1 m/s$$
Quite satisfying. We should mention here that the surface velocities
in which each of these phenomena is seen is not a single value and
varries in some range, but here, by a rough estimate,
we showed that this variations is near the same point for all of them.
And this is the first step for experimentally showing the uniqueness
of the cause.
\subsection{The Force Between The Tangent Bulks}
As the second step we compare the forces, $f_m$, exerted on the bodies
to push the bulks to eachother - In addition to surface tension that 
noramlly causes the bulks to coalesce.
For the case of drops the only force is gravity. If we estimate it
for the heaviest floating drops, with about $6mm$ diameter, we find the numerical
result: $ f_m = 300 \mu N $.

For the drops with temperature difference, the maximum force reported
is about $ 100 \mu N $.

And for the repel of currents, we can easily estimate the strike.
As we know $ f=dp/dt$. In the present situation we should
use the Newton equation in the horizontal direction, so we can write
$ f=d(mv_x)/dt$. By considering a hard collision we can rewrite
the equation like this: $f= 2v_x\, dm/dt$. With another rough estimate
we can write $ \frac {dm}{dt}=\rho S v_y $ where $S$ is the cross section
of the currents. This gives the last equation:
$$ f_m=\rho \frac{\pi}{4} d^2 v_x v_y $$
where $d$ is the diameter and $(v_x, v_y)$ is the speed of current just
at the point of collision. This time the answer for the limit
is about $ 400 \mu N $.

Another time we see a good agreement between the parameters of different
effects, and we can pick another step forward to the experimental
evidence of the unique physics lying behind these phenomena.

{\it I specially thank A. Shariati for his grateful help in preparing the article
and P. Forughi and S. Siavoshi for their help in the 
observations. I also thank M.R. Ejtehadi and N. Rivier for helpful comments.
I'm grateful to M. Hafezi and M. Sedighi for recording the experiments 
and preparing the pictures with patience. 
}


\begin{thebibliography}{99}
\bibitem{peyman} My friend "Peyman Khorsand" showed it to me for the first time.
\end{thebibliography}
\end{document}